\documentclass[conference]{IEEEtran}
\IEEEoverridecommandlockouts

\usepackage{cite}
\usepackage{amsmath,amssymb,amsfonts}
\usepackage{graphicx}
\usepackage{textcomp}
\usepackage{xcolor}
\usepackage{tikz}
\usepackage{pgfplots}
\pgfplotsset{compat=1.18}
\usepackage{subcaption}

\newcommand{\sigmaalpha}{\sigma/\alpha}
\tikzstyle{lns-aod-style}   =[no marks, line join=round, color=black!30!cyan,   dashed, thick]
\tikzstyle{jpl04-aod-style} =[no marks, line join=round, color=black!10!orange, solid, thick]

\pgfplotsset{paxisstyle/.style={%
    width=\linewidth,
    height=0.22\paperheight,
    grid=major,
    grid style={dashed,gray!30},
    xlabel=shadowing variance over path loss coefficient $\sigmaalpha$,
    ylabel=success rate $p_\mathrm{succ}$,
    xmin=-0.01, xmax=7.01,
    ymin=-0.01, ymax=1.01,
    xtick={0.0,1.0,...,7.0},
    ytick={0.0,0.1,...,1.1},
    x tick label style={anchor=north},
    legend style={nodes={scale=0.65, transform shape}, draw=white!10!black,
                  legend cell align=left},
    legend pos=north east,
}}

\newcommand{\Sn}{S_n}
\newcommand{\Pp}[1]{P_{#1}}
\newcommand{\Fhat}{\hat{f}}
\newcommand{\irrep}[1]{[#1]}
\newcommand{\noiselevel}{\ensuremath{\sigma/\alpha}}
\newcommand{\Pind}[1]{\widetilde{P}_{#1}}
\newcommand{\Qp}{Q_{\mathrm{pos}}}
\newcommand{\Kp}{K_{\mathrm{pos}}}
\DeclareMathOperator{\range}{ran}
\DeclareMathOperator{\vecop}{vec}
\newcommand{\vc}[1]{\vecop(#1)}

\usepackage{amsthm}
\theoremstyle{plain}
\newtheorem{thm}{Theorem}
\theoremstyle{remark}
\newtheorem{rem}{Remark}

\begin{document}

\newcommand{\copyrightnotice}{%
\begin{tikzpicture}[remember picture, overlay]
  \node[anchor=south, yshift=-0.62in, text width=\textwidth, align=center, font=\footnotesize]
    at (current page.north)
    {\hyphenpenalty=10000\exhyphenpenalty=10000 This paper is a preprint (IEEE accepted status). IEEE copyright notice. \copyright~2026 IEEE. Personal use of this material is permitted. Permission from IEEE must be obtained for all other uses, in any current or future media, including reprinting/republishing this material for advertising or promotional purposes, creating new collective works, for resale or redistribution to servers or lists, or reuse of any copyrighted component of this work in other works.\par};
\end{tikzpicture}}

\title{\copyrightnotice Wireless Node Ranking via Fourier-Driven Maximum-Likelihood Estimation}

\author{%
\IEEEauthorblockN{Daniel Schneider}
\IEEEauthorblockA{\textit{Independent} \\
Germany \\
mail.schneiderd@gmx.de}
\and
\IEEEauthorblockN{Hannes Frey}
\IEEEauthorblockA{\textit{University of Koblenz} \\
Koblenz, Germany \\
frey@uni-koblenz.de}
}

\maketitle

\begin{abstract}
We apply non-abelian Fourier theory to the problem of maximum-likelihood estimation of the order of a line topology, based on signal strength measurements under shadowing only. This yields up to a 60-fold speedup over standard methods for uncorrelated shadowing.

Moreover, for correlated Gaussian shadowing with a singular covariance (e.g.\ the joint path-loss model), maximum-likelihood recovery is exact and independent of the shadowing variance (an infinite Le Cam threshold), which our implementation confirms for $\sigma/\alpha$ as high as 7.0. 
The key object is the covariance's null space. Its dimension is fixed by the topology; its orientation depends on the correlation function only through the per-link variances.

Although the mechanism formally requires an exact line, it is robust to small perturbations of the formation: an approximate kernel recovers as well as the exact one. We explore the dependence on the perturbation in detail for five nodes.

\end{abstract}


\section{Introduction}
Wireless sensor and ad-hoc networks are frequently deployed in physically elongated, near-linear arrangements --- along roads, pipelines, tunnels, building corridors, or the rows of a photovoltaic plant~\cite{abdellatif2016} --- where the \emph{order} of the nodes along the line is itself the quantity of interest (for routing, in-order data fusion, or fault localization), while GPS or anchor infrastructure is unavailable or too costly~\cite{lotker2004}. Received signal strength (RSSI), by contrast, is reported by virtually every radio at no extra cost. \emph{Wireless node ranking} is the problem of recovering the order of such nodes from signal-strength measurements alone.

Even in the simple case of a line topology with uniform spacing, the problem is hard once one insists on \emph{best possible} solutions --- by which we mean the maximum-likelihood estimator (MLE). While there are heuristics like \emph{order derivation} (AOD) \cite{zhu2011,zhu2013} in the wireless literature, the least-squares method (ALS) we proposed in \cite{FS21} (also cf. \cite{SF23}) is NP-hard in general. Its benefit is, as we will see in this work, that its solution is the MLE in the case of log-normal shadowing (LNS), i.e. uncorrelated Gaussian shadowing with fixed variance $\sigma^2$. As we have pointed out in \cite{SF23}, AOD can dominate ALS when correlation is present. However, AOD is only a heuristic when its underlying assumption does not hold, as is the case, for instance, in the joint path loss models (\cite{agrawal2009}, \cite{SF23}). Thus it is natural to ask for the MLE also in the case of correlated shadowing.

What comes as a surprise is that the concept of node ranking is known as \emph{seriation} in the mathematical statistics community with --- to the best of our knowledge --- no connection to the wireless community.  In the words of Atkins et al. \cite{atkins1998}, seriation is the \emph{sequencing of elements in such a way that highly correlated pairs of elements are near each other.} This can clearly be solved by finding the maximizer of the graph correlation, which is the way Kondor states the \emph{quadratic assignment problem} (QAP) in his seminal work \cite{kondor09} --- which in turn we tackle with the Fourier Framework he developed there based on the Clausen Fourier Transform \cite{clausen1989}.
Even more surprising is that the noise-free seriation case is solved by a tool the wireless community knows well --- spectral graph theory, with the Fiedler eigenvector at its center~\cite{atkins1998} --- which nonetheless never led it to seriation. We adopt these seriation methods (Fiedler and the minimax-optimal SABRE~\cite{issartel2024}) as polynomial-time benchmarks. Crucially, these are spectral or convex \emph{relaxations} built for polynomial-time, minimax-rate recovery; none computes the maximum-likelihood estimator, even under i.i.d.\ Gaussian noise --- where our LNS algorithm, solving the exact least-squares QAP, \emph{is} the MLE and already outperforms them. The decisive gap is under \emph{correlated} shadowing: there the MLE is \emph{best possible} --- $100\%$ recovery --- which the i.i.d.-built seriation methods cannot reach.

\emph{Contributions.} We give a maximum-likelihood treatment of node ranking. Under uncorrelated LNS the MLE is the QAP, which we solve exactly with a non-abelian Fourier branch-and-bound (\emph{Fourier B\&B}) built on the Kondor bound, up to $60\times$ faster than the Gilmore--Lawler bound (GLB). Under correlated shadowing with a \emph{singular} covariance, we prove the MLE is determined entirely by the covariance \emph{kernel} and recovers the order exactly, at every noise level (Theorem~\ref{thm:kernel-mle}), which we confirm up to $\sigma/\alpha=7$. Finally, we investigate how the kernel depends on perturbations both on the topology and the covariance matrix directly. 

\section{Related Work}\label{sec:related}
Several works address ordering or localization in linear wireless networks without
coordinate infrastructure.
Lotker et al.~\cite{lotker2004} introduced range-free ranking via connectivity alone, and
Zhu et al.~\cite{zhu2013} proposed the order-derivation algorithm (AOD), which
greedily assigns positions by assuming each node's strongest neighbor lies
in the direction of increasing index; this assumption fails at high $\sigma/\alpha$ under uncorrelated noise,
while spatial correlation in fact reinforces it~\cite{SF23}.
RSS-based sequencing has further been validated experimentally in buildings~\cite{ergin2014,ergin2018}
and addressed with distributed and self-configuration heuristics for dense
linear deployments~\cite{komguem2017,abdellatif2016}.
None of these works frame the problem as maximum-likelihood estimation or analyze
performance under correlated shadowing.

Gudmundson~\cite{gudmundson1991} proposed the canonical exponential autocorrelation
model $e^{-\Delta/d_0}$ for shadow fading over a spatial separation $\Delta$.
Agrawal and Patwari~\cite{agrawal2009} introduced the joint path-loss model (JPLM),
deriving inter-link shadowing correlations via a line-integral over a Gaussian
random field, with empirical correlations up to 0.33 in multi-hop networks.
We use the variance-scaled variant (vsJPLM) introduced in~\cite{SF23} to fix
the link-length-dependent variance of the original model.

The ordering problem is equivalent to a QAP~\cite{SF23,loiola2007}: maximizing the graph correlation \emph{is} the QAP~\cite{kondor09}.
The Gilmore--Lawler bound~\cite{gilmore1962} is the standard branch-and-bound lower
bound; the Fourier B\&B we use replaces it with the tighter Kondor bound~\cite{kondor09}.

The seriation literature (see Section~\ref{sec:background}) studies exactly our problem
under iid noise. Fogel et al.~\cite{fogel2013} reformulate seriation as the 2-SUM QAP
and derive convex (doubly-stochastic QP) relaxations for noisy recovery.
Issartel et al.~\cite{issartel2024} achieve the minimax-optimal rate
$O((\sigma/\gamma)\sqrt{\log n / n})$ under iid sub-Gaussian noise, with $\gamma$ their
signal-strength parameter.
Natik and Smith~\cite{natik2021} prove that Fiedler seriation is consistent in the
noisy graphon model, though at a suboptimal rate compared to SABRE.
Both works assume independent noise entries.
The regime of our correlation models lies entirely outside the scope of existing seriation theory.

\section{Theoretical Background}\label{sec:background}
\subsection{Seriation: Fiedler and SABRE}\label{sec:robinson}
In 1951, archaeologist W.S. Robinson introduced a matrix condition to seriate pottery types across excavation strata into a temporal sequence \cite{robinson1951}: a symmetric $F \in \mathbb{R}^{n \times n}$ is a \emph{Robinson matrix} if its entries decrease monotonically away from the diagonal, i.e.\ $F_{jk} < F_{ik}$ whenever $k \le i < j$ and $F_{ik} < F_{jk}$ whenever $i < j \le k$.
\emph{Seriation} is the problem of recovering a hidden permutation $\pi \in S_n,$ where $S_n$ is the \emph{symmetric group}, from a noisy observation $A = F_\pi + N$, where $F_\pi$ denotes a Robinson $F$ permuted by $\pi$ and $N$ is a noise matrix.

The noise-free case is solved by the Fiedler vector: for the graph Laplacian $L=D-A$ of
the similarity matrix $A$, with $D=\mathrm{diag}(A\mathbf{1})$, the eigenvector $v_2$ of
the second-smallest eigenvalue $\lambda_2(L)$ is strictly monotone along the true ordering
whenever $A$ is exactly Robinson ($N=0$)~\cite{atkins1998}, so sorting the nodes by its
entries recovers $\pi$ exactly. This is \emph{Fiedler seriation}.

As the noise $N$ grows the Fiedler vector degrades. SABRE (\emph{Seriate by Aggregating
Bisections and by Re-Evaluating},~\cite{issartel2024}) is the minimax-optimal estimator for
the noisy regime and, under correlated shadowing, our strongest polynomial-time baseline. Its
central object, playing the role the Fiedler vector plays above, is a \emph{debiased} row
distance $\widehat D_{ij}$: replacing the noise-inflated self-products of the row inner
products by a robust near-neighbor surrogate cancels the noise-variance bias, so
$\widehat D_{ij}$ consistently estimates the latent inter-node distance. SABRE then
recovers $\pi$ by aggregating local left/right bisections along these distances.

SABRE targets a normalized latent model ($O(1)$ entries, known $\sigma$), so our log-distance RSSI matrix $A_{ij}=-10\,\alpha\log_{10}|i-j|+N_{ij}$ needs adapting. Its debiased-distance stage --- the source of the noise robustness --- carries over unchanged; its bisection-and-refinement stages do not, since their thresholds are tuned to the normalized model and the log-distance signal bends the row geometry into an \emph{arch} (horseshoe) on which any distance-embedding folds. We therefore order the nodes arch-resistantly --- from the line-boundary node (least total similarity), a greedy nearest-neighbor path through the debiased distances, refined by 2-opt on the 2-SUM cost --- which is exact noise-free.

\subsection{Non-commutative Fourier Transforms}\label{sec:fourier}
In the following, we will always assume Young's orthogonal representation (YOR) is being used, cf. \cite{KB08}.
Following \cite{kondor09}, given a weighted graph $\mathcal{G}$ on $n$ vertices with
adjacency matrix $A\in\mathbb{R}^{n\times n}$, define its \emph{graph function}
$f_A\colon S_n\to\mathbb{R}$ by $f_A(g) = A_{g(n),\,g(n-1)}$.
The function $f$ is constant on each left $S_{n-2}$-coset $g S_{n-2}$ (the $n(n-1)$ blocks $S_{i,j}=\{g\in
S_n\mid g(n)=i,\,g(n-1)=j\}$), which is the key structural property exploited by the Fourier machinery.

Further, define the \emph{graph correlation} of two symmetric $n\times n$ matrices $A$ and $B$ under permutation $\pi\in S_n$ as $f(\pi) = \sum_{i,j} A_{ij}\,B_{\pi(i),\pi(j)} = \mathrm{tr}(A\,P_\pi B P_\pi^\top)$,
where $P_\pi$ is the permutation matrix of $\pi$.
Finding $\hat\pi=\arg\max_\pi f(\pi)$ is the \emph{quadratic assignment problem} (QAP).
The following bound limits the maximum of $f$ over $S_n$ in terms of the order-2 Fourier coefficients ($\lambda = [n], [n-1,1], [n-2,2], [n-2,1,1]$) of the non-commutative \emph{Fourier transform} $\hat{f}(\lambda)=\sum_{\pi \in S_n} f(\pi)\,\rho_\lambda(\pi)$, where $\rho_\lambda, \lambda \vdash n,$ are the irreducible representations, and a linear assignment problem $\mathrm{LAP}(F):=\max_{\pi\in\Sn}\mathrm{tr}(\Pp{\pi}F)$, solved by the Hungarian algorithm in $O(n^3)$.
The \emph{refined Kondor bound} is
\begin{align}\label{eq:bound513}
	B_{\mathrm{kond}}(f) &= \frac{1}{n!}\Bigl[(n{-}1)\mathrm{LAP}(F)-(n{-}2)\Fhat(\irrep{n})\notag\\
	&\quad{}+\sum_{\lambda\in\{\irrep{n{-}2,2},\,\irrep{n{-}2,1,1}\}} d_\lambda\,\|\Fhat(\lambda)\|_1\Bigr],
\end{align}
where $d_\lambda$ is the dimension of $\rho_\lambda$, $\|\cdot\|_1$ the trace norm, and $F_{ij}=\sum_{\pi:\,\pi(j)=i} f(\pi)$ the \emph{first-order marginal matrix}. Eq. \eqref{eq:bound513} is our refinement of the bound (5.13) in~\cite{kondor09}; the derivation is given in the extended version~\cite{tech26}. We have $\max_{\pi\in\Sn} f(\pi) \le  B_{\mathrm{kond}}(f)$.

The Fourier transform $\hat{f}_\lambda$ is computed efficiently via the Clausen
decomposition~\cite{KB08,clausen1989}.
For each coset representative $g\in S_n/S_{n-2}$, define the coset
function $h_g\colon S_{n-2}\to\mathbb{R}$ by $h_g(\tau)=f(g\tau)$.
Then $\hat{f}(\lambda) = \sum_{g\in S_n/S_{n-2}}\rho_\lambda(g)\,\bigoplus_{\lambda^-}[\hat{h}_g]_{\lambda^-}$ \cite{KB08}, where $\lambda^-$ ranges over Young diagrams obtained from $\lambda$ by
removing one box and $\bigoplus$ denotes a direct sum of Fourier blocks. This
reduces the Fourier transform over $S_n$ to
$n(n-1)$ transforms over $S_{n-2}$, giving the Clausen-type fast Fourier
transform on $S_n$~\cite{clausen1989}.

If $f$ is a graph function, then $h_g$ is constant which leads to the fact that $\hat{f}_\lambda$ is non-zero in YOR only for $\lambda = (n), (n-1,1), (n-2,2)$ and $(n-2,1,1)$, i.e. $f$ is of order 2 and thus \emph{band-limited}.
Now $f$ is itself not a graph function, but can be written as sum of graph functions \cite{kondor09}, $f(g) = \tfrac{1}{(n-2)!}\sum_{\pi\in S_n} f_A(g\pi)\,f_B(\pi)$, which by the convolution theorem factors in Fourier space as $\hat{f}(\lambda) = \tfrac{1}{(n-2)!}\,\hat{f}_A(\lambda)\,(\hat{f}_B(\lambda))^\top$ for $\lambda\vdash n$.
Together with the band-limitedness of graph functions, this means
$\hat{f}(\lambda)=0$ unless $\lambda\in\{(n),(n{-}1,1),(n{-}2,2),(n{-}2,1,1)\}$.
This structure is what makes the (refined) Kondor bound~\eqref{eq:bound513} computable in $O(n^3)$ at each node of the Fourier B\&B~\cite{kondor09}.
Moreover, the bound works particularly well in our node ranking context, pruning many of the inferior branches in the low to mid-noise range.

\subsection{Correlated shadowing models}\label{sec:shadowing}
The shadowing enters the observation model $A=P_{\pi^\star}BP_{\pi^\star}^\top+N$ of
Section~\ref{sec:mle} as the additive noise $N$, with $N_{ij}$ the log-normal shadowing
on the link between the nodes at positions $i$ and $j$. Fast fading is not modeled: it
averages out over the RSSI samples collected per link~\cite{SF23}. We recall the spatially
correlated models of~\cite{SF23,agrawal2009} in this matrix notation.

Correlation arises because nearby links cross overlapping terrain. We resolve it onto
the $n{-}1$ unit segments of the line: the segments carry a shadowing vector
$\xi=(\xi_1,\dots,\xi_{n-1})^\top$ with covariance $C$, and a link's shadowing is the sum
of the segments it spans, up to a fixed per-link scale, $N=D\,E\,\xi$ with
$E\in\{0,1\}^{m\times(n-1)}$ ($m=\binom n2$) the link--segment incidence matrix and $D$ a
positive diagonal.
The shadowing is thus a \emph{singular} Gaussian, $N\sim\mathcal N(0,K)$ with
$K=D\,E\,C\,E^\top D$ and $\operatorname{rank}K=n-1<m$ for non-singular $C$. Its kernel
collects the linear relations satisfied identically by \emph{every} realization of $N$:
for $D=I$ the additive, or \emph{betweenness}, relations $N_{ik}=N_{ij}+N_{jk}$, in
general their $D$-weighted counterparts. Its dimension is
$\binom n2-(n-1)=\tfrac{(n-1)(n-2)}2$ either way, and this kernel is what the
maximum-likelihood estimator of Section~\ref{sec:mle} exploits.

The segment covariance $C$ and the scaling $D$ fix the model. The joint path-loss model
(JPLM, \cite{agrawal2009}) obtains $\xi$ from a line integral of an isotropic Gaussian
field, so $\mathrm{Cov}(\xi_k,\xi_l)$ decays with segment separation over a correlation
length $\delta$; we report results against $d=d_{12}/\delta$, the neighbor spacing
$d_{12}$ in units of $\delta$. Because the original JPLM ($D=I$) gives links of different
length different variances, we follow~\cite{SF23} and use the \emph{variance-scaled} JPLM
(vsJPLM), which rescales every link to a common variance $\sigma^2$
($\operatorname{diag}K=\sigma^2$) and so keeps the constant-variance property of
log-normal shadowing even when $d\gg1$; this scaling is $D$, and it trades plain
additivity for its weighted form.
Uncorrelated log-normal shadowing (LNS), our baseline, is
\emph{not} a member of this family: its $K=\sigma^2 I$ has full rank, hence no kernel and
no noise-free subspace.

\section{Maximum Likelihood Estimation}\label{sec:mle}
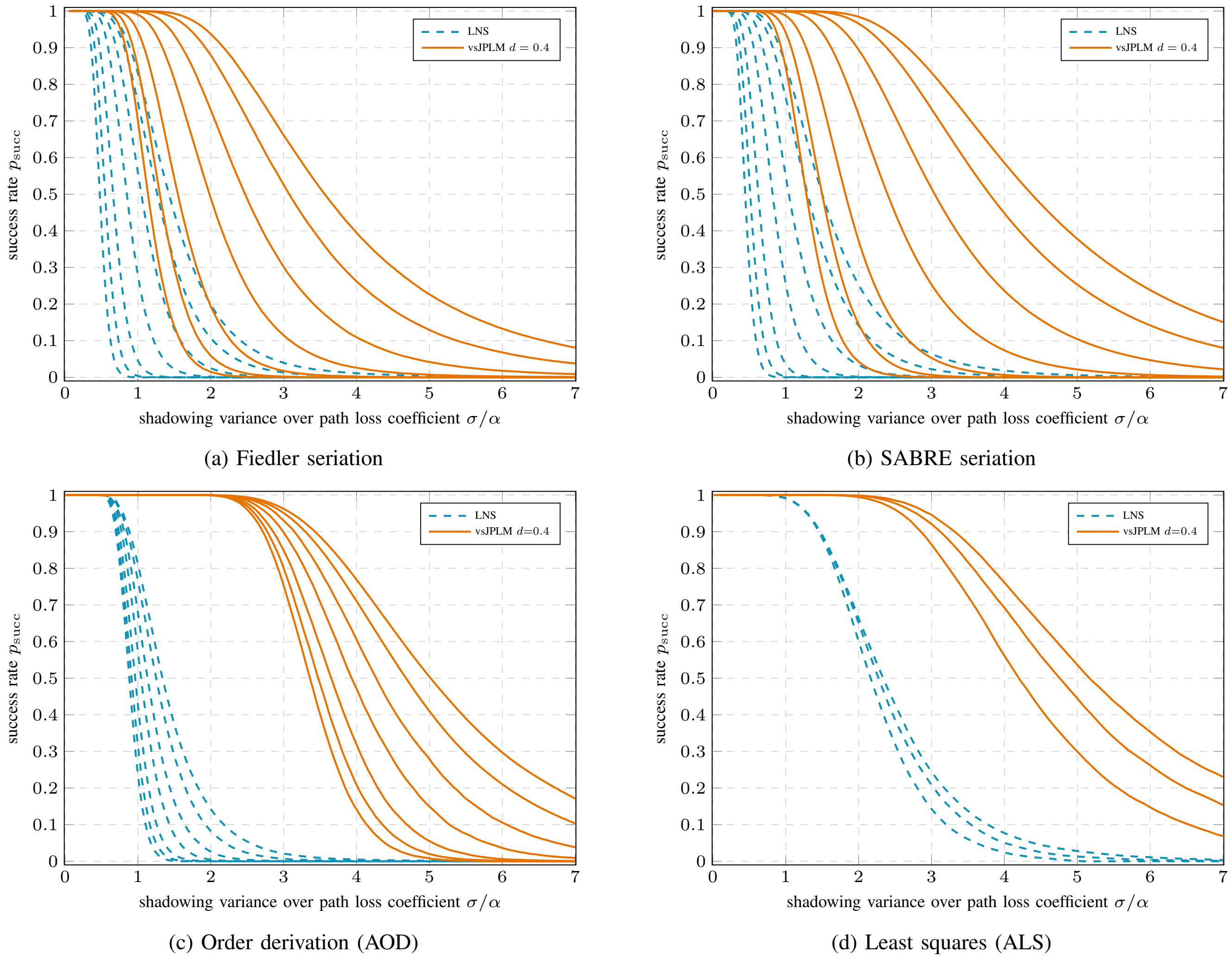
\begin{figure*}[tb]
\centering
\begin{subfigure}[t]{0.49\textwidth}\centering
\begin{tikzpicture}[scale=1.0]
\begin{scriptsize}
\begin{axis}[paxisstyle, height=0.2475\paperheight,
             xmin=-0.01, xmax=7.01, xtick={0.0,1.0,...,7.0}]
  \addplot[lns-aod-style] table[x=r, y expr=\thisrow{F_exact}/\thisrow{N},
      col sep=comma] {data/n9_lns_N200000.csv};
  \addlegendentry{LNS}
  \addplot[lns-aod-style, forget plot] table[x=r,
      y expr=\thisrow{F_exact}/\thisrow{N}, col sep=comma]
    {data/n11_lns_N200000.csv};
  \addplot[lns-aod-style, forget plot] table[x=r,
      y expr=\thisrow{F_exact}/\thisrow{N}, col sep=comma]
    {data/n15_lns_N200000.csv};
  \addplot[lns-aod-style, forget plot] table[x=r,
      y expr=\thisrow{F_exact}/\thisrow{N}, col sep=comma]
    {data/n21_lns_N200000.csv};
  \addplot[lns-aod-style, forget plot] table[x=r,
      y expr=\thisrow{F_exact}/\thisrow{N}, col sep=comma]
    {data/n31_lns_N200000.csv};
  \addplot[lns-aod-style, forget plot] table[x=r,
      y expr=\thisrow{F_exact}/\thisrow{N}, col sep=comma]
    {data/n41_lns_N200000.csv};
  \addplot[lns-aod-style, forget plot] table[x=r,
      y expr=\thisrow{F_exact}/\thisrow{N}, col sep=comma]
    {data/n51_lns_N200000.csv};
  \addplot[jpl04-aod-style] table[x=r, y expr=\thisrow{F_exact}/\thisrow{N},
      col sep=comma] {data/n9_d0.4_N200000.csv};
  \addlegendentry{vsJPLM $d=0.4$}
  \addplot[jpl04-aod-style, forget plot] table[x=r,
      y expr=\thisrow{F_exact}/\thisrow{N}, col sep=comma]
    {data/n11_d0.4_N200000.csv};
  \addplot[jpl04-aod-style, forget plot] table[x=r,
      y expr=\thisrow{F_exact}/\thisrow{N}, col sep=comma]
    {data/n15_d0.4_N200000.csv};
  \addplot[jpl04-aod-style, forget plot] table[x=r,
      y expr=\thisrow{F_exact}/\thisrow{N}, col sep=comma]
    {data/n21_d0.4_N200000.csv};
  \addplot[jpl04-aod-style, forget plot] table[x=r,
      y expr=\thisrow{F_exact}/\thisrow{N}, col sep=comma]
    {data/n31_d0.4_N200000.csv};
  \addplot[jpl04-aod-style, forget plot] table[x=r,
      y expr=\thisrow{F_exact}/\thisrow{N}, col sep=comma]
    {data/n41_d0.4_N200000.csv};
  \addplot[jpl04-aod-style, forget plot] table[x=r,
      y expr=\thisrow{F_exact}/\thisrow{N}, col sep=comma]
    {data/n51_d0.4_N200000.csv};
\end{axis}
\end{scriptsize}
\end{tikzpicture}
\subcaption{Fiedler seriation}\label{fig:fiedler-sweep}
\end{subfigure}\hfill
\begin{subfigure}[t]{0.49\textwidth}\centering
\begin{tikzpicture}[scale=1.0]
\begin{scriptsize}
\begin{axis}[paxisstyle, height=0.2475\paperheight,
             xmin=-0.01, xmax=7.01, xtick={0.0,1.0,...,7.0}]
  \addplot[lns-aod-style] table[x=r, y expr=\thisrow{SABRE_exact}/\thisrow{N},
      col sep=comma] {data/sabre_n9_lns.csv};
  \addlegendentry{LNS}
  \addplot[lns-aod-style, forget plot] table[x=r, y expr=\thisrow{SABRE_exact}/\thisrow{N},
      col sep=comma] {data/sabre_n11_lns.csv};
  \addplot[lns-aod-style, forget plot] table[x=r, y expr=\thisrow{SABRE_exact}/\thisrow{N},
      col sep=comma] {data/sabre_n15_lns.csv};
  \addplot[lns-aod-style, forget plot] table[x=r, y expr=\thisrow{SABRE_exact}/\thisrow{N},
      col sep=comma] {data/sabre_n21_lns.csv};
  \addplot[lns-aod-style, forget plot] table[x=r, y expr=\thisrow{SABRE_exact}/\thisrow{N},
      col sep=comma] {data/sabre_n31_lns.csv};
  \addplot[lns-aod-style, forget plot] table[x=r, y expr=\thisrow{SABRE_exact}/\thisrow{N},
      col sep=comma] {data/sabre_n41_lns.csv};
  \addplot[lns-aod-style, forget plot] table[x=r, y expr=\thisrow{SABRE_exact}/\thisrow{N},
      col sep=comma] {data/sabre_n51_lns.csv};
  \addplot[jpl04-aod-style] table[x=r, y expr=\thisrow{SABRE_exact}/\thisrow{N},
      col sep=comma] {data/sabre_n9_d0.4.csv};
  \addlegendentry{vsJPLM $d=0.4$}
  \addplot[jpl04-aod-style, forget plot] table[x=r, y expr=\thisrow{SABRE_exact}/\thisrow{N},
      col sep=comma] {data/sabre_n11_d0.4.csv};
  \addplot[jpl04-aod-style, forget plot] table[x=r, y expr=\thisrow{SABRE_exact}/\thisrow{N},
      col sep=comma] {data/sabre_n15_d0.4.csv};
  \addplot[jpl04-aod-style, forget plot] table[x=r, y expr=\thisrow{SABRE_exact}/\thisrow{N},
      col sep=comma] {data/sabre_n21_d0.4.csv};
  \addplot[jpl04-aod-style, forget plot] table[x=r, y expr=\thisrow{SABRE_exact}/\thisrow{N},
      col sep=comma] {data/sabre_n31_d0.4.csv};
  \addplot[jpl04-aod-style, forget plot] table[x=r, y expr=\thisrow{SABRE_exact}/\thisrow{N},
      col sep=comma] {data/sabre_n41_d0.4.csv};
  \addplot[jpl04-aod-style, forget plot] table[x=r, y expr=\thisrow{SABRE_exact}/\thisrow{N},
      col sep=comma] {data/sabre_n51_d0.4.csv};
  \end{axis}
\end{scriptsize}
\end{tikzpicture}
\subcaption{SABRE seriation}\label{fig:sabre-sweep}
\end{subfigure}

\vspace{1ex}

\begin{subfigure}[t]{0.49\textwidth}\centering
\begin{tikzpicture}[scale=1.0]
\begin{scriptsize}
\begin{axis}[paxisstyle, height=0.2475\paperheight]
  \addplot[lns-aod-style] table[x=r, y expr=\thisrow{AOD_exact}/\thisrow{N},
      col sep=comma] {data/n9_lns_N100000.csv};
  \addlegendentry{LNS}
  \addplot[lns-aod-style, forget plot] table[x=r,
      y expr=\thisrow{AOD_exact}/\thisrow{N}, col sep=comma]
    {data/n11_lns_N100000.csv};
  \addplot[lns-aod-style, forget plot] table[x=r,
      y expr=\thisrow{AOD_exact}/\thisrow{N}, col sep=comma]
    {data/n15_lns_N50000.csv};
  \addplot[lns-aod-style, forget plot] table[x=r,
      y expr=\thisrow{AOD_exact}/\thisrow{N}, col sep=comma]
    {data/n21_lns_N50000.csv};
  \addplot[lns-aod-style, forget plot] table[x=r,
      y expr=\thisrow{AOD_exact}/\thisrow{N}, col sep=comma]
    {data/n31_lns_N50000.csv};
  \addplot[lns-aod-style, forget plot] table[x=r,
      y expr=\thisrow{AOD_exact}/\thisrow{N}, col sep=comma]
    {data/n41_lns_N50000.csv};
  \addplot[lns-aod-style, forget plot] table[x=r,
      y expr=\thisrow{AOD_exact}/\thisrow{N}, col sep=comma]
    {data/n51_lns_N50000.csv};
  \addplot[jpl04-aod-style] table[x=r, y=p, col sep=comma] {data/aod_n9_d0.4.csv};
  \addlegendentry{vsJPLM $d{=}0.4$}
  \addplot[jpl04-aod-style, forget plot] table[x=r, y=p, col sep=comma] {data/aod_n11_d0.4.csv};
  \addplot[jpl04-aod-style, forget plot] table[x=r, y=p, col sep=comma] {data/aod_n15_d0.4.csv};
  \addplot[jpl04-aod-style, forget plot] table[x=r, y=p, col sep=comma] {data/aod_n21_d0.4.csv};
  \addplot[jpl04-aod-style, forget plot] table[x=r, y=p, col sep=comma] {data/aod_n31_d0.4.csv};
  \addplot[jpl04-aod-style, forget plot] table[x=r, y=p, col sep=comma] {data/aod_n41_d0.4.csv};
  \addplot[jpl04-aod-style, forget plot] table[x=r, y=p, col sep=comma] {data/aod_n51_d0.4.csv};
\end{axis}
\end{scriptsize}
\end{tikzpicture}
\subcaption{Order derivation (AOD)}\label{fig:aod-sweep}
\end{subfigure}\hfill
\begin{subfigure}[t]{0.49\textwidth}\centering
\begin{tikzpicture}[scale=1.0]
\begin{scriptsize}
\begin{axis}[paxisstyle, height=0.2475\paperheight]
  \addplot[lns-aod-style]
      table[x=r, y=p, col sep=comma] {data/bnb_n9_lns.csv};
  \addlegendentry{LNS}
  \addplot[lns-aod-style, forget plot]
      table[x=r, y=p, col sep=comma] {data/bnb_n11_lns.csv};
  \addplot[lns-aod-style, forget plot]
      table[x=r, y=p, col sep=comma] {data/bnb_n15_lns.csv};
  \addplot[jpl04-aod-style]
      table[x=r, y=p, col sep=comma] {data/bnb_n9_d0.4.csv};
  \addlegendentry{vsJPLM $d{=}0.4$}
  \addplot[jpl04-aod-style, forget plot]
      table[x=r, y=p, col sep=comma] {data/bnb_n11_d0.4.csv};
  \addplot[jpl04-aod-style, forget plot]
      table[x=r, y=p, col sep=comma] {data/bnb_n15_d0.4.csv};
\end{axis}
\end{scriptsize}
\end{tikzpicture}
\subcaption{Least squares (ALS)}\label{fig:bnb-sweep}
\end{subfigure}
\caption{Success rate $p_\mathrm{succ}$ vs.\ shadowing level $\sigmaalpha$ under LNS (cyan) and vsJPLM $d{=}0.4$ (orange). Node counts (from top down) $n\in\{9,11,15,21,31,41,51\}$, except ALS with $n\in\{9,11,15\}$.}
\label{fig:recovery-grid}
\end{figure*}

We now derive the maximum-likelihood estimator of the ordering $\pi^\star$ from the
observation $A=P_{\pi^\star}BP_{\pi^\star}^\top+N$, for both uncorrelated and correlated
Gaussian shadowing. Under uncorrelated shadowing the MLE reduces to the
least-squares QAP of~\cite{FS21,SF23}, which we re-establish and solve exactly with the
Fourier B\&B of Section~\ref{sec:fourier}. Under correlated shadowing this is
no longer optimal --- that AOD can beat ALS at moderate $\sigma/\alpha$ already shows the
least-squares estimator is misspecified there. The singular covariance of the
correlated models, however, makes the correlated MLE not merely better but \emph{exact}:
its likelihood is confined to the affine subspace fixed by $\ker K$, and projecting onto
that kernel recovers $\pi^\star$ up to reversal at every noise level (Theorem~\ref{thm:kernel-mle}),
which our experiments confirm at $100\%$ success at $\sigma/\alpha=7$. 
\subsection{Link model}
Following our previous work~\cite{SF23}, the $n$ nodes lie on a line in an unknown order; write
$v_1,\dots,v_n$ for the nodes in their true spatial sequence, with equal neighbor
spacing $d_{12}$, so nodes at positions $i$ and $j$ are at distance $|i-j|\,d_{12}$. Each
transmission is attenuated by distance-dependent path loss and by shadowing, so the
received power (in dB) on the link between them is
\begin{equation}\label{eq:rxpower}
  -10\,\alpha\log_{10}\!\big(|i-j|\,d_{12}\big)+\mathrm{const}+N_{ij},
\end{equation}
with $\alpha$ the path-loss exponent and $N_{ij}$ the shadowing of
Section~\ref{sec:shadowing}. Collecting the measurements into a symmetric similarity
matrix and absorbing the constant yields the observation model
$A=P_{\pi^\star}BP_{\pi^\star}^\top+N$ with $B_{ij}=-10\,\alpha\log_{10}|i-j|$,
where $B$ is the noise-free path-loss template of the ordered line, $\pi^\star$ the
unknown permutation taking measured indices to line positions, and $N\sim\mathcal N(0,K)$
the shadowing. As $B$ is fixed and strictly Robinson (Section~\ref{sec:robinson}), recovering
$\pi^\star$ from $A$ is exactly the seriation/QAP of Section~\ref{sec:fourier}; the rest
of this section derives its maximum-likelihood solution.

\subsection{Least squares}
Under uncorrelated log-normal shadowing, link errors are i.i.d.\ Gaussian in the dB domain: $N_{ij}=N_{ji}\stackrel{\text{iid}}{\sim}\mathcal{N}(0,\sigma^2)$ for $i<j$, with zero diagonal. Independence on the upper triangle factorizes the joint density, giving
\begin{equation*}
\ell(\pi) \;=\; \mathrm{const} \;-\; \frac{1}{2\sigma^2}\sum_{i<j}\!\bigl(A_{ij}-(\Pp{\pi}B\Pp{\pi}^\top)_{ij}\bigr)^{2}.
\end{equation*}
Since $\sigma^2$ does not depend on $\pi$, the MLE is the least-squares estimator
$\hat\pi_{\text{MLE}}=\arg\min_\pi \|A-\Pp{\pi}B\Pp{\pi}^\top\|_F^2$.
Using $\|\Pp{\pi}B\Pp{\pi}^\top\|_F^2=\|B\|_F^2$ and that $\|A\|_F^2$ is constant in $\pi$,
\begin{equation*}
\|A-\Pp{\pi}B\Pp{\pi}^\top\|_F^2 \;=\; \|A\|_F^2+\|B\|_F^2 - 2\,\mathrm{tr}\!\bigl(\Pp{\pi}^\top A\,\Pp{\pi}B\bigr),
\end{equation*}
so $\hat\pi_{\text{MLE}}=\arg\max_\pi \mathrm{tr}(\Pp{\pi}^\top A\,\Pp{\pi}B)$, which is precisely the QAP objective $f(\pi)$ solved exactly by the Fourier B\&B.

\subsection{Correlated shadowing: the singular-covariance MLE}\label{sec:singmle}
We vectorize the $m=\binom n2$ upper-triangular entries: $a=\vc{A}$, $b=\vc{B}$, and
$\mu_\pi=\vc{\Pp{\pi}B\Pp{\pi}^\top}=\Pind{\pi}\,b$, where $\Pind{\pi}$ is the permutation
that $\pi$ induces on the $m$ links, so the model reads $a=\mu_{\pi^\star}+N$ with
$N\sim\mathcal N(0,K)$ \emph{singular} of rank $n-1<m$. We sketch the consequences here
and refer to~\cite{tech26} for the derivation.

The rank deficiency, not the particular $K$, is the whole mechanism. Every realization of
$N$ lies in the column space $\range(K)$, so the orthogonal projector
$Q=I-KK^{+}$ onto the complementary \emph{noise-free} subspace
$\ker K$, of dimension $\tfrac{(n-1)(n-2)}2$, annihilates the noise identically: $Q\,N=0$
for every realization.
A singular normal is supported on the affine flat $\mu_\pi+\range(K)$, so its likelihood
vanishes unless $Q(a-\mu_\pi)=0$: the MLE is feasibility followed by a generalized
least-squares fit. But the residual is \emph{exactly} zero at the truth --- its mean part
trivially, its noise part by $Q\,N=0$ --- so the fit is vacuous, the minimizer is
independent of the norm, and the MLE collapses to the $Q$-projection estimator
\begin{equation}\label{eq:qproj}
  \hat\pi=\operatorname*{arg\,min}_\pi\big\|Q(\pi)(a-\mu_\pi)\big\|^2
         =\operatorname*{arg\,min}_\pi\big\|\Qp\,(\Pind{\pi}^\top a-b)\big\|^2 .
\end{equation}
The second form uses a single canonical kernel: with $\Kp$ the covariance and $\Qp$ its
kernel projector for the natural order $1,\dots,n$, hypothesis $\pi$ merely re-indexes the same physical shadowing,
$Q(\pi)=\Pind{\pi}\,\Qp\,\Pind{\pi}^\top$, so the covariance is never re-estimated per
hypothesis and $\pi$ acts on the data alone. Recovery is exact and independent of the
noise level (the infinite Le~Cam threshold of Remark~\ref{rem:lecam}).

Expanded against a basis $v_1,\dots,v_{(n-1)(n-2)/2}$ of $\ker\Kp=\range\Qp$, the residual
\begin{equation}\label{eq:resid}
  r_\ell(\pi)=v_\ell\cdot(\Pind{\pi}^\top a-b)
  =\sum_{i<j} v_\ell[\{i,j\}]\,a_{\{\pi(i),\pi(j)\}}-v_\ell\cdot b
\end{equation}
is manifestly second-order, a function of the pairs $\{\pi(i),\pi(j)\}$. Braces denote
unordered pairs: $m$-vectors are indexed by the links $\{i,j\}$, and
$a_{\{\pi(i),\pi(j)\}}$ is well defined by the symmetry of $A$. The problem is
thus solved by a constraint-propagation branch-and-bound, which we call the \emph{kernel
least-squares} algorithm (AKLS): the truth satisfies every $r_\ell=0$, so a partial
assignment is pruned as soon as one $r_\ell$ is bounded away from zero over its completions.

\begin{thm}\label{thm:kernel-mle}
For every singular correlated shadowing model of Section~\ref{sec:shadowing}, the residuals
\eqref{eq:resid} vanish, almost surely and at every noise level, only at the true ordering
and its reversal, so the $Q$-projection \eqref{eq:qproj} is the MLE and identifies the
line up to reversal.
\end{thm}

\noindent The proof is given in~\cite{tech26}.

\begin{rem}[infinite Le Cam threshold]\label{rem:lecam}
For every singular model of Section~\ref{sec:shadowing}, distinct orderings induce
mutually singular observation laws, so the Le~Cam two-point method returns only the
trivial bound, and by Theorem~\ref{thm:kernel-mle} the recovery threshold in
$\noiselevel$ is infinite~\cite[\S2.3, Thm.~2.2]{tsybakov2009}. This is a property of the
exact kernel: with an approximate one, recovery degrades gracefully rather than staying
exact (Fig.~\ref{fig:pert-sweep}).
\end{rem}

\section{Recovery Experiments}\label{sec:recovery}
Figure~\ref{fig:recovery-grid} is the centerpiece: it sweeps the exact-recovery rate against the noise level $\sigma/\alpha$ for all four algorithms --- Fiedler, SABRE, AOD, and ALS --- under both uncorrelated (LNS) and correlated (vsJPLM) shadowing, across a range of node counts. We then compare the exact solvers' runtimes.

The cases $n=3,5,7$ have been studied in great detail in \cite{SF23}, so we focus mainly on odd $n\ge 9$. As in \cite{SF23}, for correlated shadowing $d=6.4$ leads to worse success rates than $d=0.4$; we therefore present $d=0.4$ (very high correlation), the favorable regime that flatters AOD and Fiedler seriation. We cap ALS at $n=15$: it may run even at $n=51$ under low noise, but the runtime can explode on a bad sample, which makes a systematic evaluation at the $N_{\mathrm{runs}}=10^4$ runs we use for ALS impossible beyond $n=15$ on our hardware.

\subsection{Heuristics and ALS}
Figure~\ref{fig:recovery-grid} compares the success rates of four algorithms, three of them polynomial-time; the fourth is the least-squares estimator ALS, solved exactly by the Fourier B\&B. The two seriation algorithms have similarly shaped performance curves. Under correlated shadowing SABRE dominates Fiedler at every $n$; under LNS the two cross over near $n{=}15$, SABRE leading below it and Fiedler above. Against order derivation under LNS the picture is mixed: seriation is better at the lower node counts, AOD at $n=21,31,41,51$. ALS dominates all other algorithms under LNS for $n=9,11,15$, where its runtime is practical. Under correlated shadowing AOD is best up to $\sigmaalpha\approx4.3$, when ALS takes the lead ($n=9,11,15$ only) --- ALS is not the ML estimator there, which is why AOD can beat it.

\subsection{Exact Solver Runtime Comparison}\label{sec:solvers}
\begin{figure*}[tp]
\centering
\begin{subfigure}[t]{0.48\linewidth}
\centering
\begin{tikzpicture}[scale=1.0]
\begin{scriptsize}
\begin{axis}[
    scale only axis,
    width=0.70\linewidth, height=0.20\paperheight,
    grid=major, grid style={dashed,gray!30},
    xlabel={shadowing variance over path loss coefficient $\sigmaalpha$},
    x tick label style={anchor=north},
    xmin=-0.01, xmax=7.01, xtick={0.0,1.0,...,7.0},
    axis y line*=left,
    ylabel={mean runtime [s]},
    ymode=log, log basis y=10, ymin=4e-4, ymax=1.2e-1, ytick={1e-3,1e-2,1e-1},
    legend style={nodes={scale=0.75, transform shape}, draw=white!10!black,
                  legend cell align=left,
                  at={(axis cs:6.95,1e-2)}, anchor=south east},
  ]
  \addplot[only marks, mark=o, mark size=0.8pt, color=black!10!orange, opacity=0.35]
      table[x=r, y=mean_t_glb, col sep=comma] {data/glb_sweep_n9_mean.csv};
  \addlegendentry{GLB (Gilmore--Lawler)}
  \addlegendimage{no marks, color=black!30!purple, thick}
  \addlegendentry{GLB trend}
  \addplot[no marks, color=black!30!blue, thick]
      table[x=r, y=mean_t_fourier, col sep=comma] {data/glb_sweep_n9_mean.csv};
  \addlegendentry{Fourier B\&B}
  \addplot[no marks, color=black!30!purple, thick, forget plot]
      table[x=r, y=t_trend, col sep=comma] {data/glb_sweep_n9_glbtrend.csv};
  \addlegendimage{no marks, color=black!50!green, thick, densely dashed}
  \addlegendentry{speedup (from trend)}
\end{axis}
\begin{axis}[
    scale only axis,
    width=0.70\linewidth, height=0.20\paperheight,
    xmin=-0.01, xmax=7.01,
    axis x line=none,
    axis y line*=right,
    ylabel={speedup $t_{\mathrm{GLB}}/t_{\mathrm{B\&B}}$},
    ymin=0, ymax=62,
  ]
  \addplot[no marks, color=black!50!green, thick, densely dashed]
      table[x=r, y=speedup_trend, col sep=comma] {data/glb_sweep_n9_speedup.csv};
\end{axis}
\end{scriptsize}
\end{tikzpicture}
\subcaption{Exact-solver runtime over shadowing level $\sigmaalpha$ at $n{=}9$ under
uncorrelated (LNS, Gaussian) noise; mean over $N_{\mathrm{runs}}{=}10^4$ Fourier and $N_{\mathrm{glb}}{=}20$
GLB instances per level, log-scaled. The solid purple line is a linear trend fit to the GLB runtime; the
speedup (dashed, right, linear axis) is that trend over the moving-average-filtered Fourier B\&B runtime.}
\label{fig:solver-runtime}
\end{subfigure}%
\hfill
\begin{subfigure}[t]{0.48\linewidth}
\centering
\begin{tikzpicture}[scale=1.0]
\begin{scriptsize}
\begin{axis}[
    scale only axis, name=mainplot,
    width=0.72\linewidth, height=0.20\paperheight,
    grid=major, grid style={dashed,gray!30},
    xlabel={line-perturbation variance $v$},
    ylabel={success rate $p_\mathrm{succ}$},
    xmin=0, xmax=0.0485, ymin=0.6, ymax=1.01,
    scaled x ticks=false, xtick={0,0.01,0.02,0.03,0.04,0.05},
    x tick label style={/pgf/number format/fixed, /pgf/number format/precision=2},
    legend style={nodes={scale=0.7, transform shape}, draw=white!10!black,
                  legend cell align=left, at={(axis cs:0.0535,1.0)}, anchor=north east},
  ]
  \addplot[no marks, color=black, very thick] table[x=v,y=exactK,col sep=comma]{data/pert_sweep_n5.csv};
  \addlegendentry{exact $K$ (nominal)}
  \addplot[no marks, color=black!30!blue, thick] table[x=v,y=Knom_tr8,col sep=comma]{data/pert_sweep_n5.csv};
  \addlegendentry{$\hat K$(nominal), $s{=}8$}
  \addplot[no marks, color=black!30!blue, densely dashed] table[x=v,y=Knom_tr4,col sep=comma]{data/pert_sweep_n5.csv};
  \addlegendentry{$\hat K$(nominal), $s{=}4$}
  \addplot[no marks, color=black!20!red, thick] table[x=v,y=Kpert_tr8,col sep=comma]{data/pert_sweep_n5.csv};
  \addlegendentry{$\hat K$(perturbed), $s{=}8$}
  \addplot[no marks, color=black!20!red, densely dashed] table[x=v,y=Kpert_tr4,col sep=comma]{data/pert_sweep_n5.csv};
  \addlegendentry{$\hat K$(perturbed), $s{=}4$}
  \draw[gray!55!black, thin] (axis cs:0,0.95) rectangle (axis cs:0.007,1.003);
  \coordinate (zrA) at (axis cs:0,0.95);
  \coordinate (zrB) at (axis cs:0.007,0.95);
\end{axis}
\begin{axis}[
    name=zoom, scale only axis,
    at={(mainplot.south west)}, anchor=south west, xshift=0.105\linewidth, yshift=0.022\paperheight,
    width=0.30\linewidth, height=0.0278\paperheight,
    xmin=0, xmax=0.007, ymin=0.95, ymax=1.003,
    axis background/.style={fill=white}, axis on top,
    tick label style={font=\tiny}, scaled x ticks=false,
    xtick={0,0.0035,0.007}, ytick={0.95,1.0},
    x tick label style={/pgf/number format/fixed,/pgf/number format/precision=3},
    grid=major, grid style={dashed,gray!20},
  ]
    \addplot[no marks, color=black, very thick] table[x=v,y=exactK,col sep=comma]{data/pert_sweep_n5.csv};
    \addplot[no marks, color=black!30!blue, thick] table[x=v,y=Knom_tr8,col sep=comma]{data/pert_sweep_n5.csv};
    \addplot[no marks, color=black!30!blue, densely dashed] table[x=v,y=Knom_tr4,col sep=comma]{data/pert_sweep_n5.csv};
    \addplot[no marks, color=black!20!red, thick] table[x=v,y=Kpert_tr8,col sep=comma]{data/pert_sweep_n5.csv};
    \addplot[no marks, color=black!20!red, densely dashed] table[x=v,y=Kpert_tr4,col sep=comma]{data/pert_sweep_n5.csv};
\end{axis}
\draw[gray!55!black, thin] (zrA) -- (zoom.north west);
\draw[gray!55!black, thin] (zrB) -- (zoom.north east);
\end{scriptsize}
\end{tikzpicture}
\subcaption{Recovery rate vs.\ line-perturbation variance $v$ ($n{=}5$, vsJPLM
$d{=}0.4$, $\sigmaalpha=7$): the exact singular kernel of the nominal line against approximate kernels $\hat K$ sampled from $s\in\{4,8\}$
realizations, on the nominal line and on the perturbed array, whose full-rank covariance
admits no exact kernel.}
\label{fig:pert-sweep}
\end{subfigure}
\caption{Left: exact-solver runtime at $n{=}9$ under LNS. Right: kernel sample
efficiency under line perturbation ($n{=}5$, vsJPLM).}
\label{fig:runtime-and-pert}
\end{figure*}
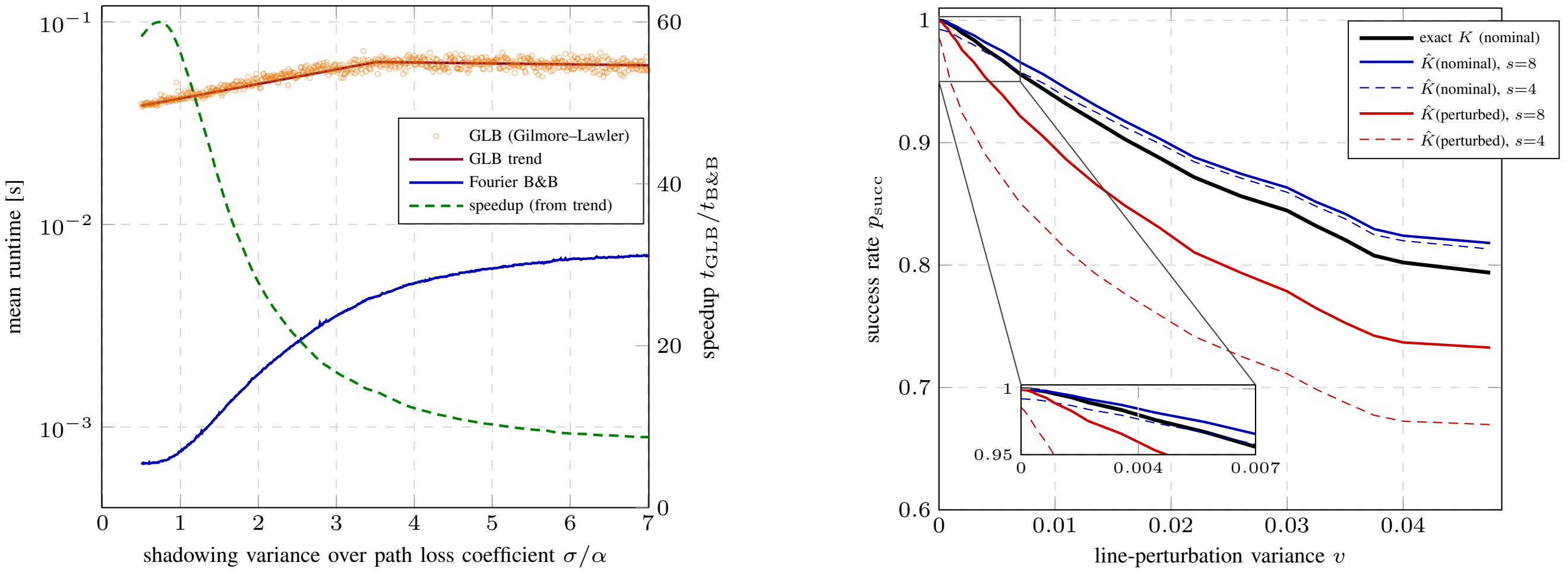

At matched $(n,\noiselevel)$ the Fourier B\&B is up to $60\times$ faster than the GLB solver~\cite{gilmore1962} at $n{=}9$ (Fig.~\ref{fig:solver-runtime}). Both runtimes grow with noise, but the Fourier B\&B has the steeper SNR dependence (${\approx}10\times$ swing against GLB's ${\approx}1.9\times$), so the speedup $t_{\mathrm{GLB}}/t_{\mathrm{B\&B}}$ (dashed) peaks at ${\approx}60\times$ near $\sigmaalpha{=}0.75$ and decays to ${\approx}9\times$ at $\sigmaalpha{=}7.0$. The Fiedler and AOD heuristics are noise-independent at ${\approx}7\,\mu$s, AOD the fastest of all methods compared. The exact solvers agree in all but $3$ cells, where a GLB exactness defect (sub-optimal return on ${\sim}10^{-4}$ instances) invalidates the local speedup. The SDP relaxation~\cite{zhao1998} agrees on sampled instances but is intractable at this size, and its margin widens with $n$: ${\approx}10^4\times$ the Fourier B\&B already at $n{=}7$, and ${\approx}10^5\times$ at $n{=}9$, where a single solve takes ${\approx}5$--$8$\,min.

\subsection{MLE for correlated shadowing}
\subsubsection{Perturbation}
AKLS, the branch-and-bound solver of our kernel MLE, restores the order of the line topology perfectly at $\sigma/\alpha=7.0$. By Theorem~\ref{thm:kernel-mle} this recovery is exact at \emph{every} noise level; what the experiment adds is the practical fact that the solver actually \emph{terminates} and returns $\pi^\star$ up to reversal, which we have tested up to $n=15$.

A natural worry is that this $100\%$ rate is an artifact of the perfectly collinear line, on which $K$ is exactly singular. To probe it we deliberately break the line: each of the $n{=}5$ nodes is displaced off it by an independent $\mathcal{N}(0,v)$ in each coordinate, which makes the covariance full-rank and destroys the exact kernel. Figure~\ref{fig:pert-sweep} ($\sigma/\alpha=7$) plots the recovery rate against $v$ for the analytic singular kernel of the nominal line and for two kernels \emph{built} by sampling, where the near-null space is read off an empirical covariance formed from $s$ noise realizations. The exact-line assumption is therefore not essential.

\subsubsection{Model Comparison}
Since recovery depends only on the kernel $\ker K$, we compare the singular models of Section~\ref{sec:shadowing} by their kernels rather than by their full covariances; the detailed comparison is given in~\cite{tech26}. The kernel's \emph{dimension} is fixed by the collinear topology and the segment construction, not by the covariance, so it is the same for all of them; what distinguishes them is its \emph{orientation}. That kernel is $D^{-1}\mathrm{ran}(E)^\perp$, so it sees $C$ only through the per-link variances that set $D$: all un-scaled models share one kernel whatever their correlation function, and variance scaling is what rotates it.

\section{Conclusion}\label{sec:conclusion}
We treated wireless node ranking as maximum-likelihood estimation of the ordering of a line topology from RSSI measurements. Under uncorrelated log-normal shadowing the MLE is the least-squares QAP, which the non-abelian Fourier B\&B (Kondor bound) solves exactly, up to $60\times$ faster than the GLB. We also identified node ranking with \emph{seriation} --- a link between the wireless and mathematical-statistics communities that appears to be new --- and used the Fiedler and SABRE seriation estimators as polynomial-time benchmarks. The central result concerns \emph{correlated} shadowing: when the covariance is singular, as for the (variance-scaled) joint path-loss model, the MLE reduces to a projection onto the covariance kernel and recovers the ordering exactly, independently of the shadowing variance (Theorem~\ref{thm:kernel-mle}). Our solver confirms this $100\%$ recovery up to $\sigma/\alpha=7$, and an approximate, sampled kernel recovers nearly as well when the line is perturbed.

\noindent\emph{Future work.} Several directions remain open: a polynomial-time approximation of the kernel MLE, presumably pairing the projection $Q$ with a seriation heuristic (Fiedler, SABRE); a classification of the recovery kernels; field experiments on a real line of nodes; and an extension beyond the line, since a grid or other structured array built from the same segment construction likewise has a singular covariance with a nontrivial kernel.

\end{document}